\documentclass[11pt]{article}
\usepackage{IAUS215}
\usepackage{epsfig,natbib}
\newcommand{\msun}{{\, {\rm M}_\odot}}
\newcommand{\n}{{\mem{n}}}
\newcommand{\lei}{{^{1}\mem{L}}}
\newcommand{\p}{{\mem{p}}}

\newcommand{\cdr}{{^{13}\mem{C}}}
\newcommand{\czw}{{^{12}\mem{C}}}

\newcommand{\nvi}{{^{14}\mem{N}}}

\newcommand{\ose}{{^{16}\mem{O}}}

\newcommand{\pei}{{^{31}\mem{P}}}

\newcommand{\nise}{{^{60}\mem{Ni}}}

\newcommand{\nizw}{{^{62}\mem{Ni}}}

\newcommand{\gdr}{{^{63}\mem{G}}}

\newcommand{\mem}[1]{{\mathrm{ #1}}}
\newcommand{\spr}{\mbox{$s$-process}}
\newcommand{\kelv}{{\,\mathrm K}}
\newcommand{\abb}[1]{Fig.\,\ref{#1}}

\newcommand{\glt}[1]{Eq.\,(\ref{#1})}
\newcommand{\gl}[1]{Eq.\,(\ref{#1})}

\def\edcomment#1{\iffalse\marginpar{\raggedright\sl#1\/}\else\relax\fi}
\reversemarginpar

\begin{document}
\vspace*{1cm}
\title{Mixing and the $s$-process in rotating AGB stars}
 \author{Falk Herwig}
\affil{Department of Physics and Astronomy, University of Victoria,
  3800 Finnerty Rd, Victoria, BC, V8P 1A1, Canada}
 \author{Norbert Langer}
\affil{Astronomical Institute, Universiteit Utrecht, P.O. Box 80000, NL-3508 TA Utrecht, The Netherlands}
\author{Maria Lugaro}
\affil{Institute of Astronomy, University of Cambridge, Madingley Road, Cambridge CB3 0HA, United Kingdom } 

\begin{abstract}
We model the nucleosynthesis during a radiative interpulse phase of
a rotating $3\msun$ Asymptotic Giant Branch (AGB)
star. We find an enhanced production of the neutron source species $\cdr$
compared to 
non-rotating models due to shear mixing of protons and $\czw$ at
the core-envelope interface. We estimate that  the resulting total
production of heavy elements by slow neutron capture (\spr) 
is too low  to account for most
observations. This due to the fact that rotationally induced mixing
during the interpulse  
phase causes a pollution of the $\cdr$ pocket
layer with the neutron  poison $\nvi$. As a result we find a  maximum
neutron exposure of $\tau_\mem{max} = 0.04\,
\mem{ mbarn^{-1}}$ in the \spr\ 
layer of our solar metallicity model with rotation. This is about a 
factor of $5 \dots 10$  less than required to reproduce the 
observed stellar \spr\ abundance patterns.
We compare our results with models that include hydrodynamic
overshooting mixing, and with simple parametric models including
the combined effects of overshooting and mixing in the
interpulse. Within the parametric model a range of mixing efficiencies
during the 
interpulse phase correlates with a spread in the \spr-efficiency. Such
a spread is observed in AGB and post-AGB stars as well as in pre-solar
SiC grains.
\end{abstract}

\section{Introduction}

The \spr\ is characterized by
neutron captures that are immediately 
followed by $\beta$-decays. Unstable isotopes involved  in the \spr\
path  have
typical life times of the order of hours; at neutron densities  of
$\mem{N_n} < 10^{10}\mem{cm}^{-3}$
they decay rather than capture another neutron
 (see Busso etal., 1999, for a general review of the \spr).
Current models (Gallino etal., 1998; Goriely \& Mowlavi, 2000; Lugaro
etal., 2002) favour low mass Asymptotic Giant Branch (AGB) stars as the
dominant production site for the \spr\  nuclei with mass numbers $A>90$. In evolved AGB stars  He-shell burning is periodically unstable
and generates thermal pulse (TP) cycles.
The main neutron producing reaction is
$\cdr(\alpha,n)\ose$ at 
temperatures as low as $T=9 \cdot 10^7\kelv$.
The neutrons are released under radiative conditions during
the interpulse phase. 

The necessary amount of $\cdr$ is produced 
if protons and $\czw$ can partially mix at the core-envelope
interface. This may happen after TP during the dredge-up phase during 
which a fraction of intershell material is carried into the envelope.
The required extra-mixing is still uncertain.
With the  H/$\czw$-ratio decreasing continuously from a few hundred in
the envelope to zero in the intershell both a $\cdr$ as well as a
$\nvi$ pocket forms. Herwig (2000) has described models with
hydrodynamical, diffusive overshooting that leads to a partial mixing
zone. However, stars rotate and in this paper we discuss the impact of
rotationally induced mixing in AGB star models on the \spr. A more detailed
discussion is available in Herwig etal.\ (2003).

\section{Properties of the Partial
Mixing Zone}
\label{sec:obs-pm}

\begin{figure}
\plotone{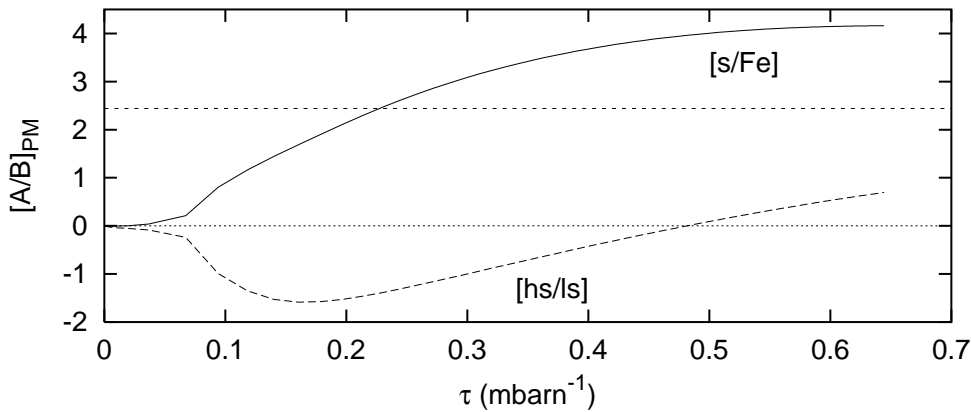} 
\caption{ \label{fig:tau-shsls} 
\spr\ indices [s/Fe] and [hs/ls]  as a function of
neutron exposure.  As explained in the text this
figure translates 
observational properties of \spr\ enriched stars into properties of
the partial mixing zone. The dashed/dotted horizontal lines are
discussed in the text. }
\end{figure}
The \spr\  abundances observed in AGB stars can be characterized by
 averaged abundance  indicators. The
index [hs/ls] describes the distribution of the \spr\ elements.
 Stars of solar metallicity show $-0.5 < \mem{[hs/ls]}
 < 0.0$, where [ls/Fe]=$\frac{1}{2}$([Y/Fe]+[Zr/Fe]) and hs is the
 average of Ba, La, Nd and Sm. The total overproduction
of \spr\ elements correlates with the index [s/Fe], which is the
 average of Y and Nd, and observationally $0< \mem{[s/Fe]} < 1$ (\mbox{Busso
 etal.,} 1995).  

The observed overproduction factors in the envelope are related to the
overproduction factors in the \spr\ zone by a number of dilution
factors, which result from the subsequent mixing events:
the He-flash convective mixing and the third
dredge-up. The envelope abundance of any species in
the envelope after dredge-up events in $m$ identical TP cycles is
related to the abundance in the \spr\ zone (partial mixing zone, PM) by
\begin{equation}
\label{math:mix}
Y_\mem{env}=q m Y_\mem{PM}\frac{M_\mem{PM}  M_\mem{DUP}}{ M_\mem{IS}
M_\mem{env}}
\end{equation}
where $M_\mem{PM}$, $M_\mem{DUP}$, $M_\mem{IS}$
and  $M_\mem{env}$ are the masses of the partial mixing zone, the 
dredged up layer, the intershell zone covered by the He-flash
convection and the envelope mass respectively. 
These quantities change with the pulse number, typical numbers in
low mass stars are  $M_\mem{DUP}
\sim 3\cdot 10^{-3}\msun$, $ M_\mem{IS}\sim 10^{-2}\msun$
and  $ M_\mem{env}\sim 0.5\msun$.  The factor $q$ describes the effect
of the overlapping of the He-flash convection zone in subsequent thermal pulses
and can be estimated to be $q=2.3$ (see Herwig etal., 2003 for
details). We may assume that not more than 20 thermal 
pulses with dredge-up contribute to the envelope enrichment with \spr\
material ($m=20$).

By evaluating \glt{math:mix} for the numbers specified above  we can derive a 
logarithmic expression that relates the average \spr\ overabundance in
the PM zone with the mass of that zone:
\begin{equation}
\label{gl:sFe-Mpm}
\log M_\mem{PM}= - \mem{[}s/Fe\mem{]}_\mem{PM} + c
\end{equation}
where $Y_\mem{env}$ in \glt{math:mix} is given by using
$\mem{[s/Fe]_{env}} = 1$, and 
$c=-0.44$ for $m=20$.

In \abb{fig:tau-shsls} we show the variation  of
$\mem{[s/Fe]}_\mem{PM}$ and $\mem{[hs/ls]_\mem{PM}}$ with the neutron
exposure in a \spr\ one-zone model. As
more neutrons are released the average \spr\ overabundance 
$\mem{[s/Fe]}_\mem{PM}$ 
increases. The partial mixing
zone should not extend into the He-shell. This puts a
lower limit on [s/Fe]$_\mem{PM}$ and thus on the neutron exposure
in the partial mixing zone. In \abb{fig:tau-shsls} the dashed
horizontal line corresponds to $M_\mem{PM}<10^{-2}\msun$ and
$c=-0.44$, and according to \gl{gl:sFe-Mpm} [s/Fe]$_\mem{PM}$ above
that line are allowed.
From [hs/ls]$_\mem{obs}<0.0$ (dotted line) it follows that the neutron
exposure should not exceed $\sim 0.45 \mem{mbarn}^{-1}$ in the partial
mixing zone. Otherwise  the envelope 
\spr\ abundance distribution will eventually be top-heavy as well, and not in
agreement with observations (Lugaro etal., 2003). 

Thus, we estimate that the neutron exposure at the end of the
interpulse phase should be in the range
$0.25 \mem{mbarn}^{-1}<\tau < 0.45 \mem{mbarn}^{-1}$. 
Using \glt{gl:sFe-Mpm} and  $\mem{[s/Fe]}_\mem{PM} =
4.0$ at the upper $\tau$ limit, the mass of the partial mixing
zone should obey $\log M_\mem{PM} > -4.44$ (assuming $m=20$).
These estimates are roughly in agreement with the 
much more detailed analysis of the
partial mixing zone by Gallino etal.\ (1998).

\section{$s$-Process nucleosynthesis in a rotating AGB model star}
\label{sec:results}
\begin{figure}[t]
\plotone{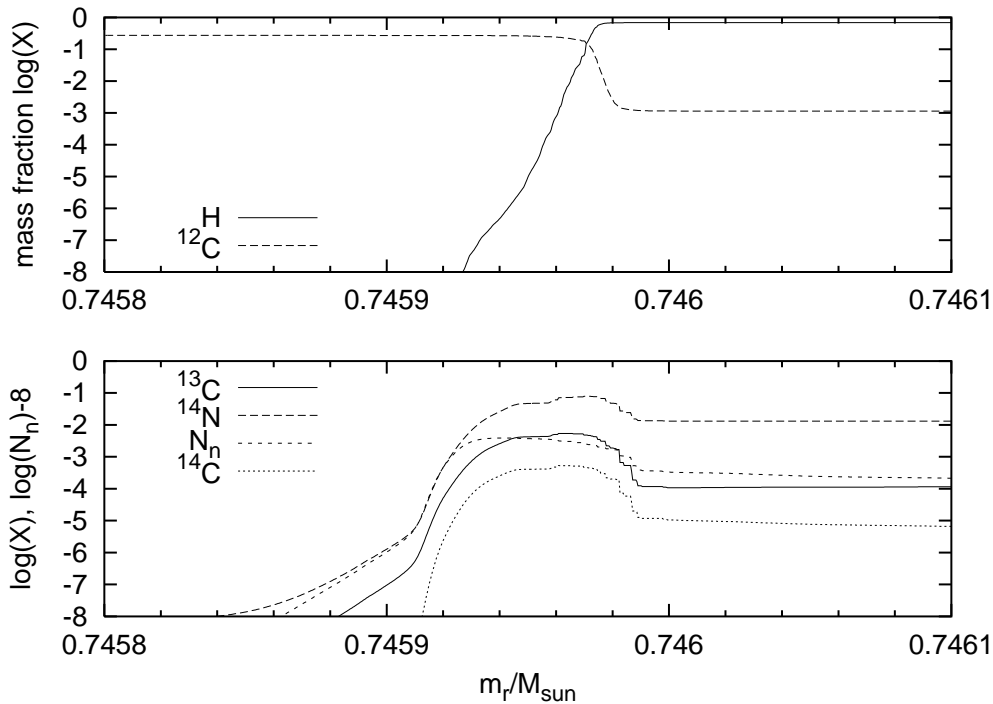} 
\caption{ \label{fig:ROT-prof} 
Abundance profiles in the partial mixing zone immediately after the
end of the dredge-up phase (upper panel) and when the
$\cdr(\alpha,\n)\ose$ reaction is activated (lower panel).
}\end{figure}
Rotation induces mixing in stars through a variety of processes
(Maeder \& Meynet, 2000). In rotating AGB stars a large
angular velocity gradient develops at the interface of the compact and fast
rotating core and the slow rotating envelope.  For our analysis we
take the corresponding mixing efficiencies and thermodynamic input
from a 3$\msun$  stellar model sequence including rotation (Langer etal.,
1999). These 
models predict that shear mixing is continuously present at the core-envelope
interface throughout the interpulse phase. For the \spr\ the effective
Lagrangian mixing efficiency is important. Note, that the partial
mixing layer is contracting significantly during the interpulse phase.
 Around the time of
$\cdr(\alpha,n)\ose$ ignition (interpulse phase $\phi \sim 0.3$), the diffusion
coefficient is $D_\mem{rot} \sim 10^{-13}\msun/\mem{yr}$, which
gives for a mixing timescale of $\sim 1000 \mem{yr}$ a mixing range of
$\sim 10^{-5}\msun$. 

We use our  \emph{nuclear
network and mixing post-processing} code to evaluate the effect
of mixing on the \spr. 
This code solves simultaneously for mixing processes
and all considered nuclear processes (charged particle reactions,
neutron captures and $\beta$-decays). In the network all relevant
charged particle reactions from hydrogen to $\pei$ have been
considered with rates mainly taken from Angulo etal.\ (1999). Neutron
captures for 15 light isotopes and 9 iron group species have been
considered and most cross sections  are from Bao etal.\ (2000). The
neutron capture on species beyond $\nise$ are considered by a two
step heavy neutron sink.  We introduce two artificial particles
($\gdr$ and $\lei$) with two extra 
reactions ($\nizw(\n,\gamma)\gdr$ and $\gdr(\n,\lei)\gdr$)  which  
together give an estimate  on \spr\ efficiency.  For the 
Maxwellian averaged neutron capture cross section of $\gdr$ we take
$\sigma_{\gdr}(8\mem{keV})=120\mem{mbarn}$, which  has been
determined from extensive one-zone model tests with a full \spr\
network.  

In \abb{fig:ROT-prof} we show the abundance profiles around the core
boundary at two times of the post-processing 
calculation of one interpulse phase of the model sequence with
rotation. The top panel shows the initial
condition  where rotationally induced mixing has caused some diffusion
of protons into the $\czw$ rich core at the end of the (in this case
shallow) third dredge-up. Goriely \& Mowlavi (2000) have shown that in such a
situation of a continuously decreasing proton abundance profile into
the $\czw$ rich core the mass range with a proton abundance of
$-2<\log X_\mem{p}<-3$ will eventually host \spr\ enriched material. 
According to this criterion, and given the required
properties of the partial  mixing zone as estimated in the previous
section the partial mixing zone obtained in this rotating model is
too thin. 

The lower panel of \abb{fig:ROT-prof} shows the situation when
neutrons are beginning to be released by the $\cdr(\alpha,n)\ose$
reaction. In models without rotation, where the
partial mixing zone results for example from hydrodynamic
overshooting, the $\nvi$  and the $\cdr$ pocket  
keep mainly separated. With rotation instead the $\nvi$ abundance exceeds that
of $\cdr$ in the entire partial mixing zone. $\nvi$ is an efficient
neutron absorber, most notably through the $(n,p)$ reaction. 
The recycling effect of the protons
released in this reaction and the subsequent $\czw(\p,\gamma)\cdr$
reaction  does not balance the poison effect of 
$\nvi$. As a result the neutron densities are low and the final
maximum neutron exposure is much lower ($\tau \sim
0.04\mem{mbarn}^{-1}$)than our estimates from the previous section
 require.   

\section{Parametric Models and Discussion}
\label{sec:discussion}
\begin{figure}[t]
\plotone{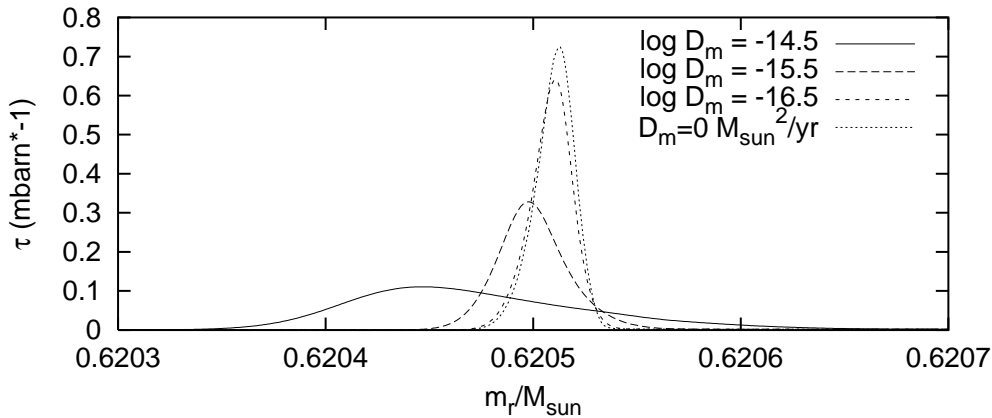} 
\caption{ \label{fig:mr-tau} 
Final neutron exposure at the end of the interpulse for different
assumptions on mixing. All cases start with a H/$\czw$ partial mixing
zone from hydrodynamic overshooting. During the interpulse no mixing
(dotted line) and constant Lagrangian mixing coefficients have been
applied ($D_\mem{m}$ in units of $\msun/\mem{yr}$).
}\end{figure}
It seems that current models of rotation in AGB stars predict
interpulse mixing coefficients that could inhibit the \spr\ during the
interpulse phase 
due to dilution of the $\cdr$ pocket and admixture of the neutron
poison $\nvi$. In order to further study the effect of mixing during the
interpulse we have constructed some parametric \spr\ models
assuming a constant diffusive mixing coefficient $D_\mem{m}$ in Lagrangian
coordinates  during the interpulse phase. We start in all
cases with abundance profiles  
at the core-envelope interface shaped by hydrodynamic overshooting as
described in Herwig (2000). The final neutron exposure at the
end of the interpulse phase is shown in \abb{fig:mr-tau} for several
cases. The case without mixing shows a larger  neutron exposure than
allowed according to the estimates in $\S$\ref{sec:obs-pm} 
For the cases with  constant $D_\mem{m}$ we find an anti-correlation
of mixing efficiency and final neutron exposure. Mixing coefficients
over a  range of two 
orders of magnitude correspond to a large spread of neutron exposure.
For larger mixing efficiencies  the pocket containing 
\spr\ enriched material becomes broader. However, the mixing
coefficients that reproduce neutron exposures estimated in
$\S$\ref{sec:obs-pm} are much
smaller than what is predicted by the full stellar evolution model
with rotation. We attempt to attribute this to unaccounted processes
of angular momentum transport like magnetic fields (see Heger
etal., these proceedings) in the AGB models. The simple
parametric models indicate  a possible solution  
for the apparent spread in \spr\ efficiency reported from pre-solar
grains (Lugaro etal., 2002a) as well as from  AGB 
and post-AGB stellar observations (Van Winckel \& Reyniers, 2000; 
Busso etal., 2001).

\paragraph{Acknowledgements:} 
F.H.\ would like to thank D.A. VandenBerg
for support through his Operating Grant from the Natural Science and
Engineering Research Council of Canada.

\end{document}